\documentclass[prd,twocolumn,floatfix,showpacs,preprintnumbers,nofootinbib]{revtex4}
\usepackage[utf8x]{inputenc}
\usepackage{mathrsfs}
\usepackage{amssymb}
\usepackage{amsmath}
\usepackage{graphicx}

\usepackage[caption = false]{subfig}
\usepackage[normalem]{ulem}
\usepackage{xcolor}
\definecolor{lcolor}{rgb}{0.5,0,0}
\definecolor{citcolor}{rgb}{0,0.3,0.0}

\usepackage[breaklinks,colorlinks,urlcolor=blue,citecolor=citcolor,linkcolor=lcolor]{hyperref}
\usepackage{mciteplus}

\newcommand{\rt}{{\mathbf{r}_T}}
\newcommand{\xt}{{\mathbf{x}_T}}
\newcommand{\bt}{{\mathbf{b}_T}}
\newcommand{\yt}{{\mathbf{y}_T}}

\newcommand{\kt}{{\mathbf{k}_T}}

\newcommand{\ptt}{p_T} % scalar
\newcommand{\ktt}{k_T} % scalar
\newcommand{\lt}{\mathbf{l}_T}

\newcommand{\ud}{\, \mathrm{d}}

\newcommand{\tr}{\, \mathrm{Tr} \, }

\newcommand{\nc}{{N_\mathrm{c}}}

\newcommand{\nr}[1]{(\ref{#1})}

\newcommand{\gev}{\ \textrm{GeV}}

\newcommand{\lqcd}{\Lambda_{\mathrm{QCD}}}
\newcommand{\as}{\alpha_{\mathrm{s}}}
\newcommand{\aem}{\alpha_{\mathrm{em}}}

\newcommand{\eq}{Eq.~}

\newcommand{\eqs}{Eqs.~}

\newcommand{\der}{\mathrm{d}}

\newcommand{\qso}{Q_\mathrm{s0}}

\begin{document}

\author{B. Duclou\'e}
\affiliation{
Institut de physique théorique, Université Paris Saclay, CEA, CNRS, F-91191 Gif-sur-Yvette, France%
}

\author{T. Lappi}
\affiliation{
Department of Physics, University of Jyv\"askyl\"a %
 P.O. Box 35, 40014 University of Jyv\"askyl\"a, Finland
}
\affiliation{
Helsinki Institute of Physics, P.O. Box 64, 00014 University of Helsinki,
Finland
}

\author{H. M\"antysaari}
\affiliation{
Physics Department, Brookhaven National Laboratory, Upton, NY 11973, USA
}

\title{
Isolated photon production in proton-nucleus collisions at forward rapidity
}


\preprint{}

\begin{abstract}
We calculate isolated photon production at forward rapidities in proton-nucleus collisions in the Color Glass Condensate framework. Our calculation uses dipole cross sections solved from the running coupling Balitsky-Kovchegov equation with an initial condition fit to deep inelastic scattering data. For comparison, we also update the results for the nuclear modification factor for pion production in the same kinematics. We present predictions for future forward RHIC and LHC measurements at $\sqrt{s_{NN}}=200$~GeV and $\sqrt{s_{NN}}=8$~TeV. 
\end{abstract}

\maketitle

%%%%%%%%%%%%%%%%%%%%%%%%%
\section{Introduction}
%%%%%%%%%%%%%%%%%%%%%%%%%%
Interpreting ultrarelativistic heavy ion collision data from the BNL-RHIC and CERN-LHC experiments requires a detailed understanding of the initial stages of the collision process. In collisions of heavy nuclei, however, the initial state effects are propagated through the space-time evolution of the produced medium, and it may be very difficult to disentangle initial state cold nuclear matter effects from final state interactions. To separately study the initial condition, or the structure of the colliding nuclei at small Bjorken $x$, the nucleus must be studied with a simpler probe. Ideally, one would want to study deep inelastic scattering, but before the future EIC or LHeC colliders are realized~\cite{Accardi:2012qut,AbelleiraFernandez:2012cc}, proton-nucleus collisions provide an environment where one does not necessarily expect the formation of a thermal medium.

Measurements of inclusive photon and hadron spectra at forward rapidities (forward being the proton-going direction) are sensitive to the small-$x$ structure of the nucleus. Even long before the LHC proton-lead results~\cite{ALICE:2012mj,Abelev:2013haa,ALICE:2012xs,Chatrchyan:2013eya,Abelev:2014dsa,Khachatryan:2015xaa}, the observed nuclear suppression for pion production at forward rapidities at RHIC~\cite{Arsene:2004ux,Adams:2006uz,Adare:2011sc} was important for nuclear parton distribution analyses and a hint of a significant suppression of the nuclear gluon distribution at small $x$~\cite{Eskola:2008ca,Eskola:2009uj}.  Recently, the possibility of using upcoming LHC isolated photon production data to constrain nuclear parton distribution functions in the collinear factorization approach, especially the gluon distribution at small $x$, has been pointed out e.g. in Ref.~\cite{Helenius:2014qla}.

At high energy (or at very small $x$), the partonic densities become very large, of the order of $1/\as$, and a convenient framework to describe QCD in this region is given by the Color Glass Condensate (CGC) effective theory~\cite{Gelis:2010nm}. The CGC formalism provides a framework to resum large logarithms of $\as \ln 1/x$ using the Balitsky-Kovchegov (BK)~\cite{Balitsky:1995ub,Kovchegov:1999yj} (or JIMWLK) evolution equations. For particle production at forward rapidities and moderate transverse momenta, these high energy logarithms can be expected to  dominate over the transverse momentum logarithms $\as \ln Q^2$ resummed by the DGLAP evolution. The leading order inclusive particle production calculations in the CGC framework have been shown to be in good agreement with a variety of RHIC and LHC data~\cite{Blaizot:2004wu,Tribedy:2010ab,Albacete:2010bs,Tribedy:2011aa,Lappi:2013zma,Ducloue:2015gfa,Ducloue:2016pqr}. Recently, there has also been significant progress in developing the theory towards NLO accuracy~\cite{Chirilli:2012jd,Ducloue:2016shw,Stasto:2013cha,Altinoluk:2014eka,Balitsky:2008zza,Lappi:2016fmu}.

In this work we calculate isolated photon and inclusive $\pi^0$ production in the kinematics relevant to upcoming proton-lead results from the LHC experiments and proton-gold and proton-aluminum processes measured at RHIC (see the RHIC Cold QCD plan~\cite{Aschenauer:2016our}). The essential ingredient in our calculation is the dipole cross section, whose evolution with~$x$ is calculated from the running coupling BK equation. The initial conditions for this evolution have been fit to HERA DIS measurements for electron-proton deep inelastic scattering in Ref.~\cite{Lappi:2013zma} and extended to nuclei using an optical Glauber procedure relying on standard nuclear geometry. 
The same dipole cross sections, without any additional parameters, have previously been used to calculate single inclusive particle production~\cite{Lappi:2013zma}, forward $J/\Psi$ production~\cite{Ducloue:2015gfa,Ducloue:2016pqr}  and Drell-Yan cross sections~\cite{Ducloue:2017zfd} in proton-proton and proton-nucleus collisions. The range of processes addressed in these works demonstrates the universality and predictive power of the dilute-dense CGC framework, combining the dipole picture of DIS with the ``hybrid'' formalism for forward hadronic collisions.  The main purpose of this paper is to extend this set of observables, all calculated consistently with the same parametrization, to photon production.

This paper is structured as follows. First, in Sec.~\ref{sec:pions} we review the leading order $\pi^0$ production cross section calculation from the CGC formalism. In Sec.~\ref{sec:photons}, we discuss how the isolated photon production cross section is calculated in the same framework. The necessary input to our calculations, the dipole-nucleus scattering amplitude, is introduced in Sec.~\ref{sec:dipole} before showing our results for RHIC and LHC in Sec.~\ref{sec:results}.

\section{Inclusive pion production}
\label{sec:pions}

Experimentally neutral pions are typically measured together with isolated photons. Thus, while two of the authors have already addressed hadron production in an earlier work~\cite{Lappi:2013zma} (see also Ref.~\cite{Albacete:2017qng}), we shall present here the results for neutral pion and isolated photon production together for an easier comparison with measurements. For this purpose let us first briefly summarize our procedure, identical to the one of Ref.~\cite{Lappi:2013zma}, for calculating identified hadron cross sections.

Inclusive pion production at forward rapidities (and at leading order) is dominated by a process where a dilute parton from the probe scatters off the strong color field of the target and fragments into a pion. 
The invariant yield in proton-nucleus collisions for $\pi^0$ production in the ``hybrid'' formalism~\cite{Dumitru:2002qt,Dumitru:2005gt,Albacete:2010bs,Tribedy:2011aa,Rezaeian:2012ye,Lappi:2013zma} is
\begin{multline}\label{eq:pizeroxs}
\frac{\der N^{pA \to \pi^0 X}(\bt)}{\der y \der^2 \kt} = \frac{1}{(2\pi)^2} \int \frac{\der z}{z^2}  \sum_i x_p f_i(x_p,\mu^2)\\
\times  S\left(\frac{\kt}{z}, \bt, x_g\right) D_{i\to \pi^0}(\mu^2,z).
\end{multline}
Here the target is probed at momentum fraction $x_g = (\ktt/z)/\sqrt{s_{NN}} e^{-y}$, and the longitudinal momentum fraction in the proton is $x_p = (\ktt/z)/\sqrt{s_{NN}} e^{y}$.
The distribution of partons $i$ in the probe is given in terms of the parton distribution function $f_i$, and $D_{i\to \pi^0}$ is the fragmentation function describing the formation of a pion out of the parton $i$. We employ a scale choice $\mu^2 = \ktt^2$ and use the CTEQ6~\cite{Pumplin:2002vw} parton distribution functions and DSS~\cite{deFlorian:2007aj} fragmentation functions at leading order in this work.

All the information about the target is encoded in the function $S(\kt, \bt, x_g)$, which describes the quark-target scattering with transverse momentum transfer $\kt$ at impact parameter $\bt$. It is obtained as the Fourier transform of a fundamental representation dipole correlator in the target color field
\begin{equation}
S(\kt, \bt, x_g) = \int \der^2 \rt e^{-i \kt \cdot \rt} S(\rt, x_g, \bt) ,
\end{equation}
with
\begin{equation}
S(\xt - \yt) = 1 - N(\xt-\yt) = \frac{1}{\nc} \langle \tr U^\dagger(\xt) U(\yt) \rangle.
\end{equation}
Here we denote by $U(\xt)$ the fundamental representation Wilson lines in the target color field, and $N$ is the dipole amplitude. The dipole correlator is obtained by fitting the HERA data in Ref.~\cite{Lappi:2013zma} and is discussed in more detail in Sec.~\ref{sec:dipole}.

The yield in Eq~\eqref{eq:pizeroxs} is calculated by summing over the parton species. In this work, we include $u,d,s$ and $c$ quarks and their antiquarks and gluons. For the gluon channel, the Wilson lines are taken in the adjoint representation, where the dipole correlator $\tilde S$ is obtained using the large $\nc$ approximation as $\tilde S = S^2$.

\section{Inclusive photon production in the CGC}
\label{sec:photons}
We consider photon production at forward rapidity, a process in which a relatively large-$x$ quark from the dilute projectile scatters off the dense gluonic target and radiates a photon, probing the target structure at small $x$. The inclusive photon yield for such a process is~\cite{Gelis:2002ki,Gelis:2002fw,JalilianMarian:2005zw,Stasto:2012ru,Dominguez:2011wm,JalilianMarian:2012bd}
\begin{multline}
\label{eq:photon_xs}
\frac{\der N^{pA \to \gamma X}(\bt)}{\der^2 \kt \der y_\gamma } = \sum_q \frac{e_q^2 \aem}{\pi (2\pi)^3}  \int \der^2 \lt \int_{x_\text{min}}\der x_p  \\
\times z^2[1+(1-z)^2]  
 \frac{q(x_p, \mu^2)}{\kt^2}  \frac{(\kt + \lt)^2}{[z\lt - (1-z)\kt]^2} \\
 \times  S(\kt+\lt, \bt, x_g).
\end{multline}
Here $z$ is the longitudinal momentum fraction of the quark carried by the photon and $\kt$ and $y_\gamma$ are the photon transverse momentum and rapidity, respectively. The quark transverse momentum $\lt$ and rapidity $y_q$ are integrated over. Here the integral over the quark rapidity is written in terms of $x_p$, the fraction of the proton momentum carried by the quark, since for a given photon kinematics the quark momentum $\lt$ and $x_p$ uniquely specify the quark rapidity. 

The lower limit for the momentum fraction $x_p$ integral is set by the photon kinematics as $x_{\text{min}} = k_T e^{y_\gamma}/\sqrt{s}$. The kinematics of the $1\to 2$ scattering is such that
\begin{align}
x_g &= \frac{|\kt| e^{-y_\gamma} + |\lt | e^{-y_q}}{\sqrt{s}} \\
x_p &= \frac{|\kt| e^{y_\gamma} + |\lt | e^{y_q}}{\sqrt{s}} \\
y_q &= \log \left( \frac{-e^{y_\gamma} |\kt| - x_q \sqrt{s}}{|\lt|} \right) \\
z &=  \frac{|\kt|}{x_p \sqrt{s}} e^{y_\gamma} .
\end{align}  
Our formalism is not applicable in the kinematics where $x_g$ in the nucleus becomes large. In practice, we approximate large-$x$ effects by freezing dipole amplitude at $x>x_0$ and set $S(\rt, x_g, \bt) = S(\rt, x_0, \bt)$ when $x>x_0$. Here $x_0$ is the initial condition for the BK evolution, as discussed in Sec.~\ref{sec:dipole}. 
Our results results for the photon nuclear suppression factor are sensitive to this domain even marginally only for $p_T >5 \gev$ in RHIC kinematics. Here the formalism is not completely applicable, and we also do not expect the experimental data to deviate from unity within the uncertainties.

As we perform a leading order calculation, we use the leading order CTEQ6 parton distribution functions~\cite{Pumplin:2002vw} to describe the quark content of the probe. In Eq.~\ref{eq:photon_xs} we include $u, d, s $ and $c$ quarks and their antiquarks with corresponding fractional electric charges $e_q$. The scale $\mu^2$ at which the PDFs are evaluated is chosen to be $\mu^2 = \max\{\lt^2, \kt^2\}$. The scale uncertainty mostly cancels in the nuclear modification factor, as we demonstrate explicitly in Appendix~\ref{appendix:scale}.

The expression for the cross section~(\ref{eq:photon_xs}) is divergent when the quark and the photon are close to each other in phase space. In particular, as discussed in Ref.~\cite{JalilianMarian:2012bd}, Eq.~\eqref{eq:photon_xs} contains a divergent contribution corresponding to $q \to \gamma$ fragmentation. In this work we are interested in  prompt photon production and do not want to include the fragmentation component. To enforce an isolation cut we multiply the integrand of Eq.~(\ref{eq:photon_xs}) by the measure function
\begin{equation}
\theta\left(\sqrt{(y_\gamma - y_q)^2 + \Delta \phi^2}-R\right).
\end{equation}
Here $\Delta \phi$ is the azimuthal angle difference between the scattered quark and the photon and $R$ a chosen isolation cone radius, which we will vary as a check of the systematics.

\section{Dipole scattering}
\label{sec:dipole}
To describe dipole-proton scattering we use the MV$^e$ parametrization from Ref.~\cite{Lappi:2013zma}. Here, the dipole-proton scattering amplitude $N=1-S$ at the initial rapidity $x_0=0.01$ is parametrized as
\begin{multline}
\label{eq:mve}
N(\rt, x=x_0) = 1 - \exp \left[ -\frac{\rt^2 \qso^2}{4} \right. \\
\left. \times \ln \left( \frac{1}{|\rt| \lqcd} + e_c \cdot e \right) \right].
\end{multline}
The impact parameter profile is assumed to factorize, and is parametrized by a constant: 
\begin{equation}\label{eq:protonbint}
\int \ud^2 \bt N(\rt, \bt, x) = \frac{\sigma_0}{2} N(\rt, x).
\end{equation}
The dipole amplitude is evolved to values of $x$ smaller than $x_0$ by solving the running coupling Balitsky-Kovchegov evolution equation.  
The parameters of the model ($\qso, e_c, \sigma_0$ and the scale of the coordinate space running coupling in the BK equation) have been obtained by fitting the HERA reduced cross section data at small $x\le 0.01$ in Ref.~\cite{Lappi:2013zma}.
The fit done in Ref.~\cite{Lappi:2013zma} includes only light quarks, but in this work we also include the charm quark contribution. As we are mostly interested in cross section ratios (namely the nuclear suppression factor $R_{pA}$), the quark mass effects should be negligible. 

The dipole-nucleus scattering amplitude is obtained by generalizing Eq.~\nr{eq:mve} at the initial condition $x=x_0$ to nuclei using an optical Glauber model (see again \cite{Lappi:2013zma}). The dipole-nucleus amplitude at $x=x_0$ is written as
\begin{multline}\label{eq:ainitc}
	N^A(\rt,\bt) = 1 - \exp\left[ -A T_A(\bt) \frac{\sigma_0}{2} \frac{\rt^2 \qso^2}{4} \right.  \\
	\left. \times \ln \left(\frac{1}{|\rt|\lqcd}+e_c \cdot e\right) \right].
\end{multline}
Here $T_A$ is the thickness function of the nucleus normalized to unity ($\int \der^2 \bt T_A(\bt)=1$).  The evolution to smaller values of $x$ is then done using the BK equation separately at each value of $b \equiv |\bt|$. We emphasize that all the other parameters besides the standard Woods-Saxon geometry that is used to determine $T_A$ are constrained by the HERA DIS data.

Here we need to calculate cross sections in the same kinematics in both proton-nucleus and proton-proton collisions. For a proton target we need to take into account the fact that the geometric size of the proton measured in deep inelastic scattering experiments, $\sigma_0/2$, is not the same as the inelastic nucleon-nucleon cross section $\sigma_\text{inel}$.  For a proton target, the cross section is obtained by integrating \eqs\nr{eq:pizeroxs} and \nr{eq:photon_xs} over the area occupied by the small-$x$ gluons in the target, which in our factorized  model for the proton impact parameter dependence yields a factor $\sigma_0/2$ as in \eq\nr{eq:protonbint}. The invariant yield reported by the experiments is defined as this cross section divided by the total inelastic cross section $\sigma_\text{inel}$. Thus, for a proton target, \eqs\nr{eq:pizeroxs} and \nr{eq:photon_xs} are effectively multiplied by $\frac{\sigma_0/2}{\sigma_\text{inel}}$, and the dipole-proton amplitude has no explicit impact parameter dependence.
For more details, see~\cite{Lappi:2013zma}. Here we use the values $\sigma_\text{inel}=42$ mb at $\sqrt{s}=200\gev$~\cite{Adams:2006uz} and $\sigma_\text{inel}=75$ mb at $\sqrt{s}=8$ TeV~\cite{Miller:2007ri}. This corresponds to a number of binary collisions $N_\text{bin}=4.948$ for p+Au collisions, $N_\text{bin}=2.568$ for p+Al collisions at RHIC and $N_\text{bin}=8.153$ for p+Pb collisions at the LHC.

\section{Results}
\label{sec:results}
Let us now present our results for the nuclear suppression factors $R_{pA}$. The advantage of this ratio compared to individual yields is that the overall normalization uncertainty can be expected to mostly cancel. We calculate 
\begin{equation}
R_{pA} = \frac{\der N^{pA}}{N_\text{bin} \der N^{pp}}.
\end{equation}
In the absence of nuclear effects, this ratio is exactly one. 

\subsection{LHC}

\begin{figure}[tb]
\centering
		\includegraphics[width=0.5\textwidth]{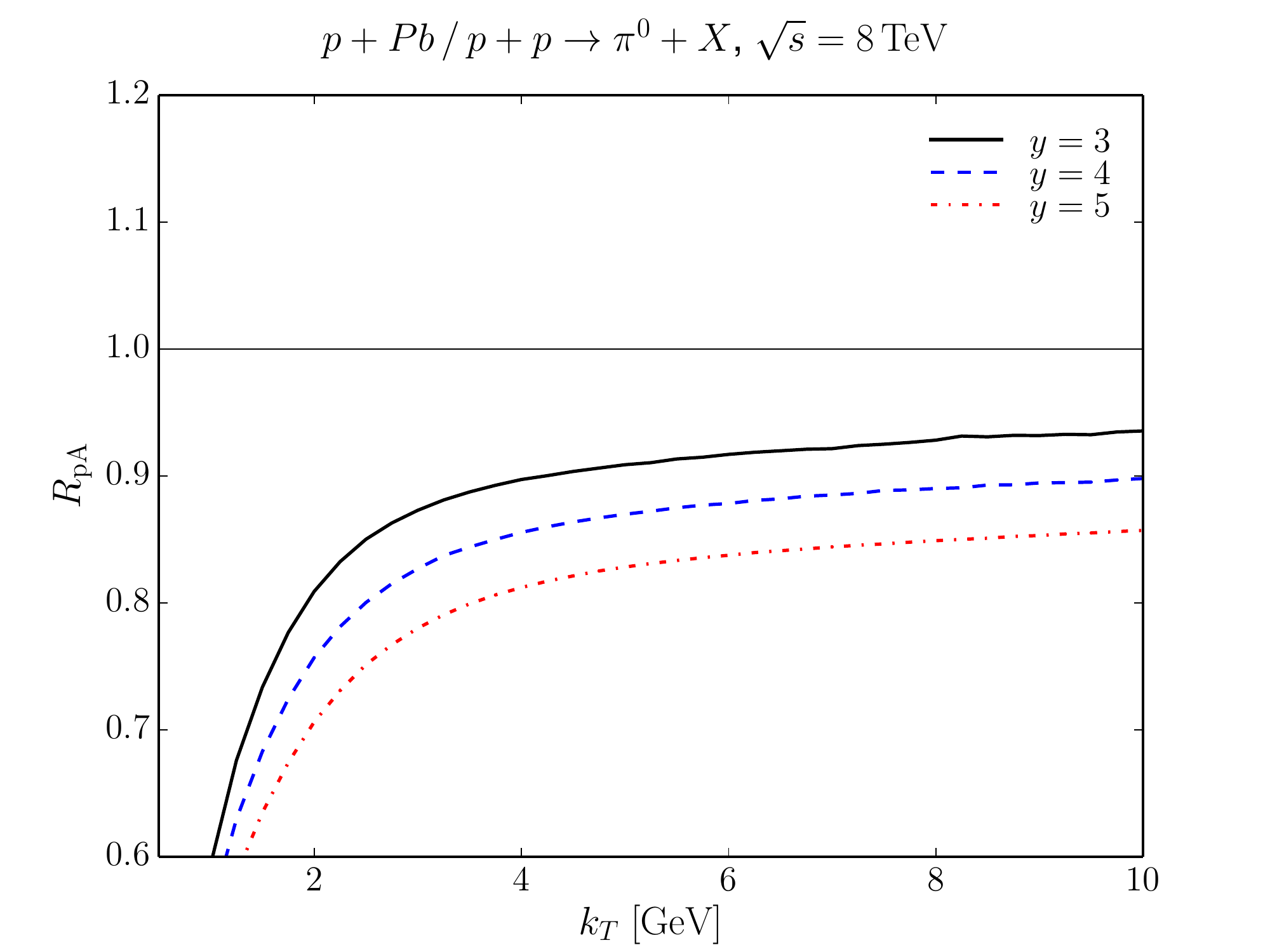} 
				\caption{Nuclear suppression factor for inclusive $\pi^0$  production at $\sqrt{s_{NN}}=8$ TeV in $p+Pb$ collisions.}
		\label{fig:rpa_pion_8tev}
\end{figure}

\begin{figure}[tb]
\centering
		\includegraphics[width=0.5\textwidth]{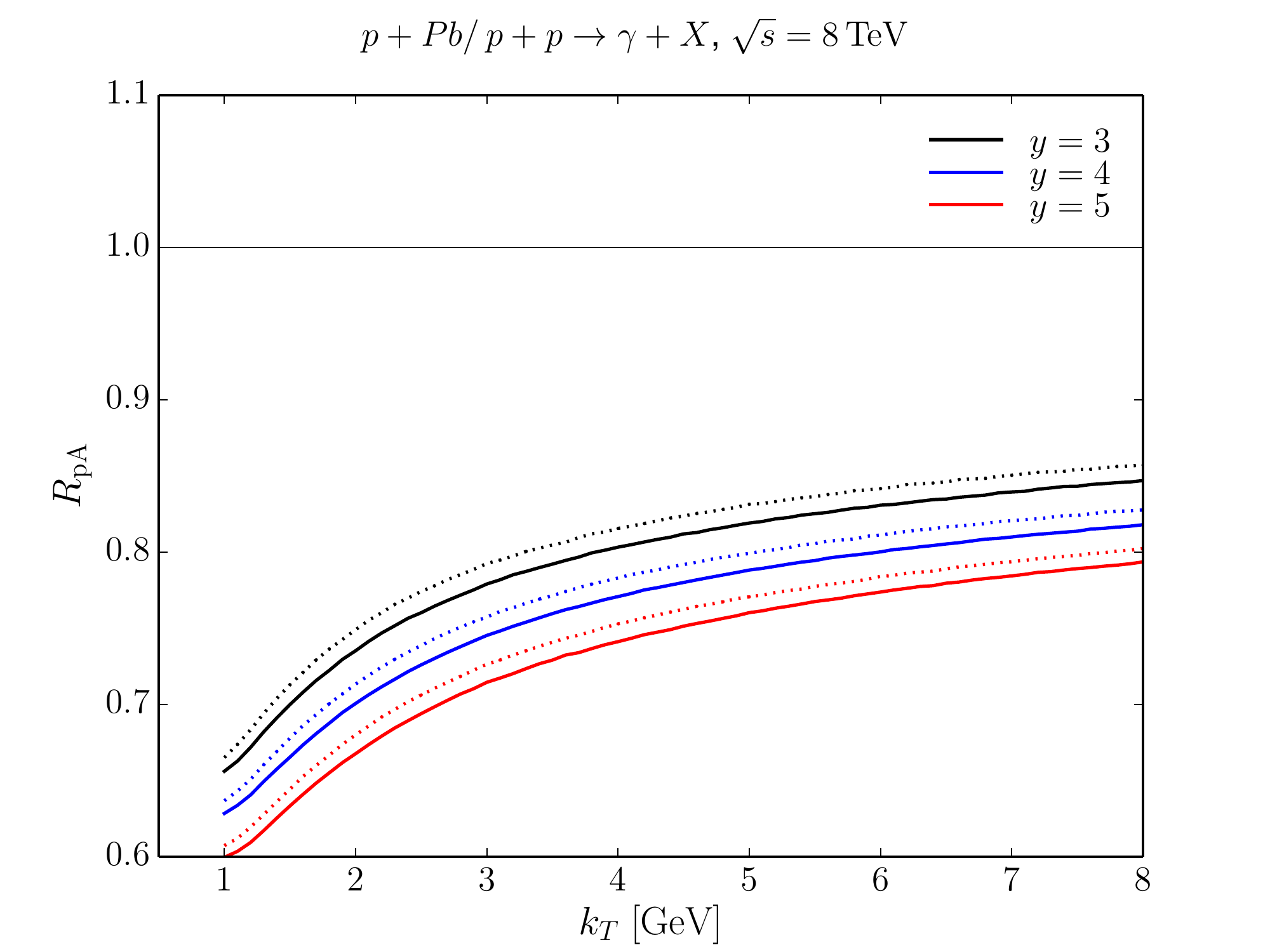} 
				\caption{Nuclear suppression factor for isolated photon production at $\sqrt{s_{NN}}=8$ TeV in $p+Pb$ collisions. Solid lines have isolation cut $R=0.4$ and dotted lines isolation cut $R=0.1$.}
		\label{fig:rpa_photon_8tev}
\end{figure}

First in Fig.~\ref{fig:rpa_pion_8tev} we present results for inclusive $\pi^0$ production at forward rapidities accessible at LHCb and, after future upgrades, also at ALICE.
The same nuclear suppression factor for isolated photon production with two different isolation cuts $R=0.4$ and $R=0.1$ is shown in Fig.~\ref{fig:rpa_photon_8tev}. Comparing the results for photon and pion production, we find that a much stronger suppression at low transverse momentum is obtained in the case of pions. The suppression factor for pions also approaches unity at high $\ktt$ faster than in the case of photons. This is expected, since in our calculation a large $\pi^0$ transverse momentum always corresponds to a large $\ktt$ in the target, leading to little nuclear modification. A large photon momentum can, on the other hand, be balanced by the recoiling quark and correspond to a small intrinsic target $\ktt$, with the associated large nuclear suppression. This pattern could change when hadron production is evaluated at NLO, where the 2-particle final state kinematics more resembles LO photon production.

The isolated photon suppression is larger than what was obtained in Ref.~\cite{Helenius:2014qla} by performing a NLO pQCD calculation with EPS09 nuclear parton distributions function~\cite{Eskola:2009uj}. Further, the suppression is expected to get stronger at more forward rapidities. This is in contrast with the calculation involving only recent nuclear PDFs, where a rapid DGLAP evolution smooths out strong nuclear effects in the gluon PDF. However, we also note that nuclear PDFs are not well constrained in the small $x$ region probed in these processes, and the corresponding predictions have large uncertainties.

The effect of different isolation cuts is also shown in Fig.~\ref{fig:rpa_photon_8tev}.  We find that $R_{pA}$ is almost insensitive to the details of the isolation procedure. A similar conclusion was made in the NLO pQCD calculation presented in Ref.~\cite{Helenius:2014qla}. Experimentally the isolation cut is defined by imposing a limit on the transverse energy in the cone, which is not possible to implement in our leading order calculation. However, the insensitivity of the nuclear suppression factor to the cone size suggests that there is relatively little uncertainty in our calculation related to the implementation of the isolation cut.

\subsection{RHIC proton-gold collisions}
After successful deuteron-gold runs where forward pion production measurements were performed~\cite{Arsene:2004ux,Adams:2006uz,Adare:2011sc}, there was a proton-gold run at $\sqrt{s_{NN}}=200$ GeV at RHIC in 2015. At forward rapidities the RHIC data is at the edge of the kinematical phase space, but it is still possible to go up to $y\sim 4$ with low $p_T$ photon and $\pi^0$ production.

\begin{figure}[tb]
\centering
		\includegraphics[width=0.5\textwidth]{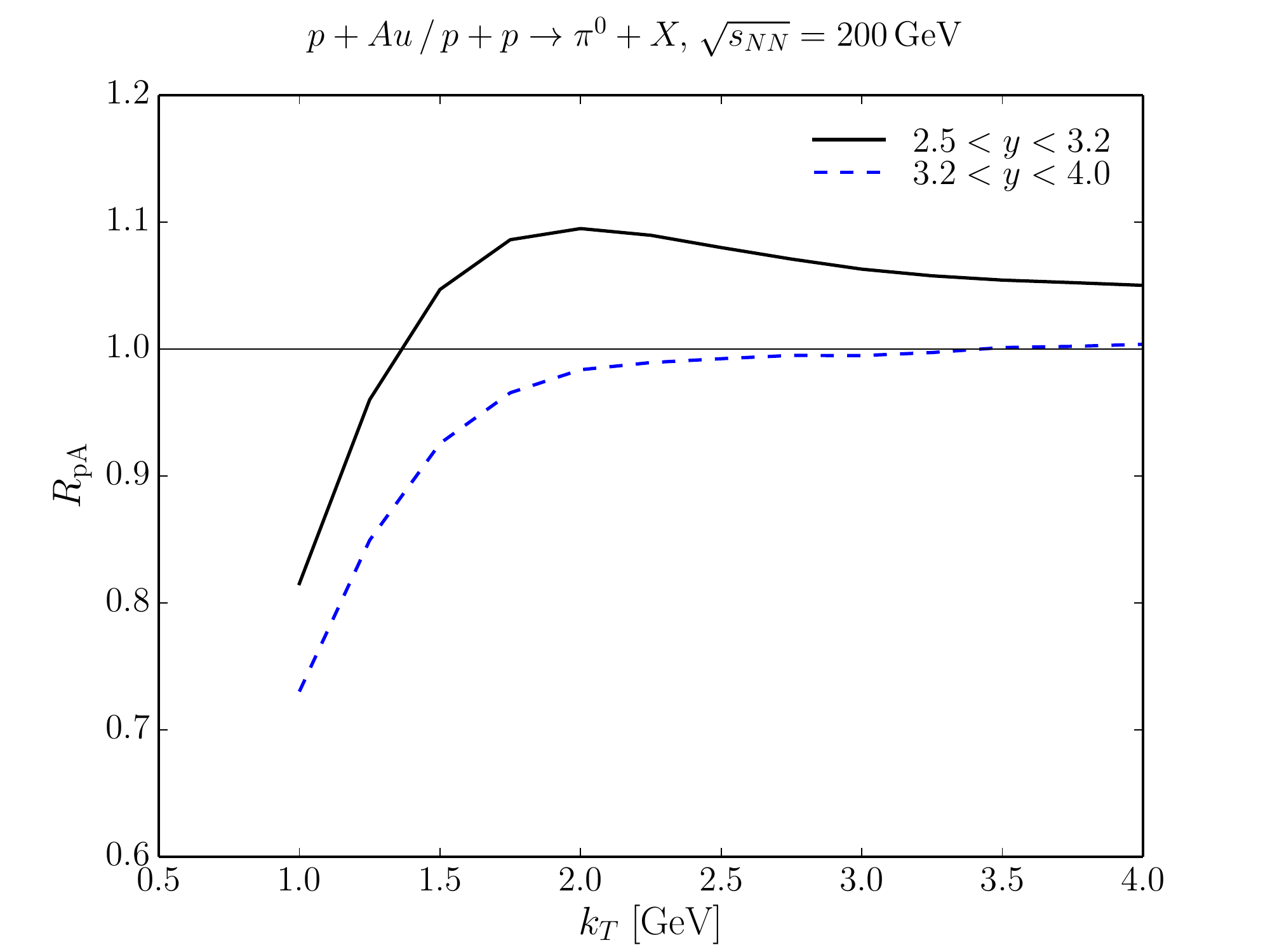} 
				\caption{Nuclear suppression factor for inclusive $\pi^0$ production in proton-gold collisions at $\sqrt{s_{NN}}=200$ GeV.}
		\label{fig:rpa_pion_200gev}
\end{figure}

Our result for the nuclear suppression factor for inclusive pion production is shown in Fig.~\ref{fig:rpa_pion_200gev} in the two rapidity bins that correspond to STAR measurements. 
As already shown by early CGC calculations~\cite{Albacete:2003iq} (see also Ref.~\cite{Kharzeev:2003wz}), a Cronin-like enhancement at low transverse momentum is visible close to the initial condition of the BK evolution. This is a result of higher saturation scales in the nucleus which makes it easier to give a transverse momentum of the order of the nuclear saturation scale to the incoming parton, compared to the parton-proton scattering in the same kinematics. This enhancement then disappears when one evolves to lower values of Bjorken $x$ (measuring pions at more forward rapidities), and the overall suppression increases as a function of rapidity.
We note that the fast decrease of $R_{pA}$ with rapidity is a feature that is also visible in the earlier BRAHMS charged hadron deuteron-gold data~\cite{Arsene:2004ux}. The values of $R_{pA}$ in our calculation are, however, larger than the ones measured by BRAHMS. Here one must note that the features of the $\ptt$ spectrum in these kinematics (such as the Cronin peak, especially in the more central rapidity bin) are very sensitive to the functional form of the dipole amplitude at the initial condition, which is rather poorly constrained by the DIS fit. 

\begin{figure}[tb]
\centering
		\includegraphics[width=0.5\textwidth]{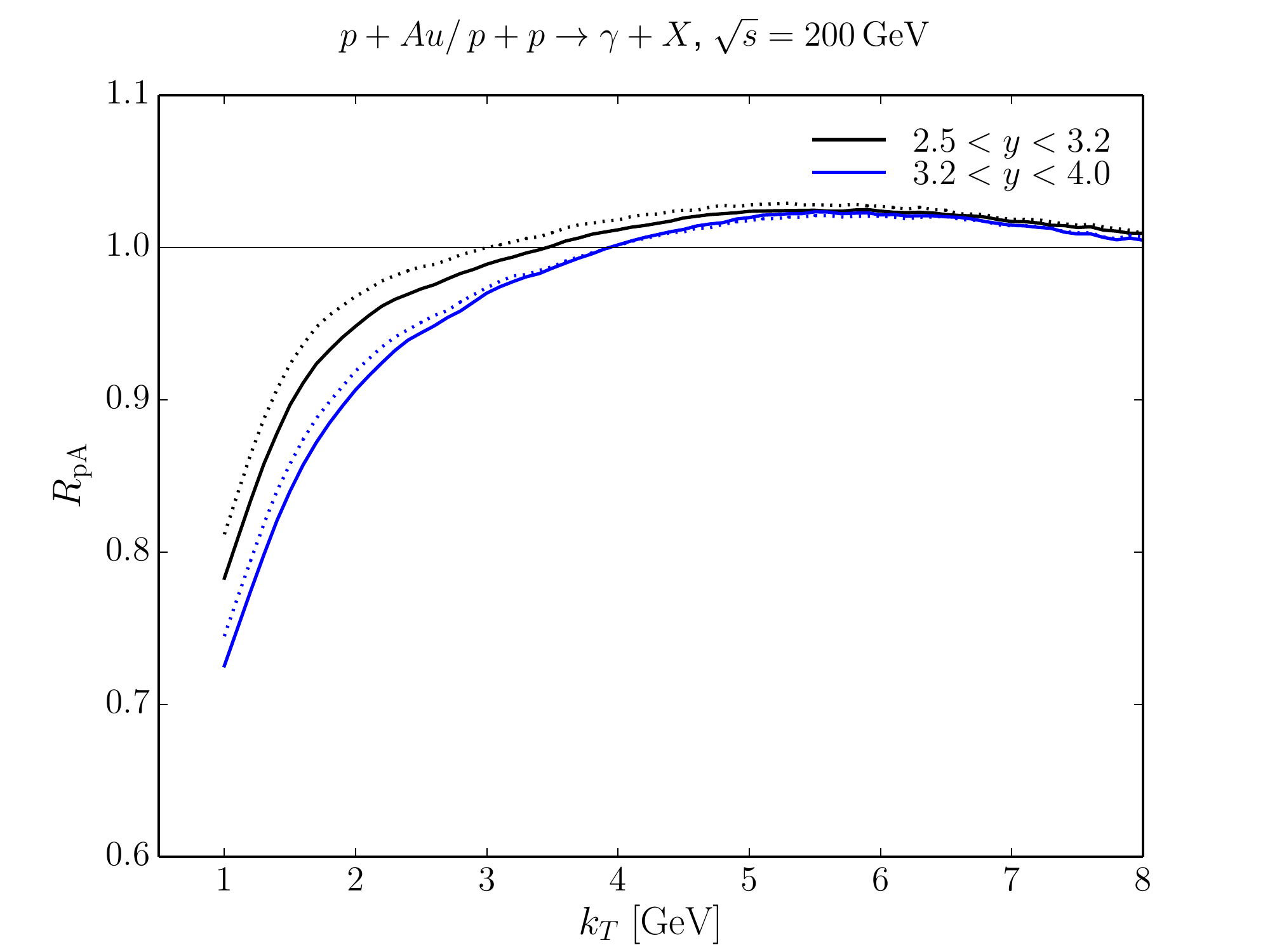} 
				\caption{Nuclear suppression factor for isolated photon production in proton-gold collisions at $\sqrt{s_{NN}}=200$ GeV. Solid lines have isolation cut $R=0.4$, and dotted lines $R=0.1$.}
		\label{fig:rpa_photon_200gev}
\end{figure}

In the same kinematics we show the nuclear suppression factor for isolated photon production in Fig.~\ref{fig:rpa_photon_200gev}. Similarly as in the case of the LHC kinematics, we find that the suppression is smaller at low $\ktt$ and has a weaker $\ktt$ dependence than for $\pi^0$ production. We also expect to see a small Cronin peak around $k_T\sim 5\gev$ in both rapidity bins. The results are very little sensitive to the details of the isolation cut, like in the LHC kinematics. 

\subsection{RHIC proton-Aluminum collisions}
The proton-aluminum collisions recorded at RHIC provide a possibility to study how nuclear effects evolve as a function of the nuclear mass number $A$ (see also Ref.~\cite{Kowalski:2007rw}). In comparison to gold with $A=197$, in the case of aluminum ($A=27$) we expect significantly smaller nuclear effects.  We note that our optical Glauber model, which uses a Woods-Saxon distribution to generalize the dipole-proton amplitude to the dipole-nucleus case (see Eq.~\eqref{eq:ainitc}), may not be accurate with such a light nucleus (see also the related discussion in \cite{Ducloue:2016pqr}). In particular, when we calculate the minimum bias cross sections, in the case of $\pi^0$ production approximately $80\%$ of the cross section comes from regions where the saturation scale of the nucleus falls below that of the proton. In that region we get a contribution that explicitly gives $R_{pA}=1$ by construction (see Ref.~\cite{Lappi:2013zma}).

\begin{figure}[tb]
\centering
		\includegraphics[width=0.5\textwidth]{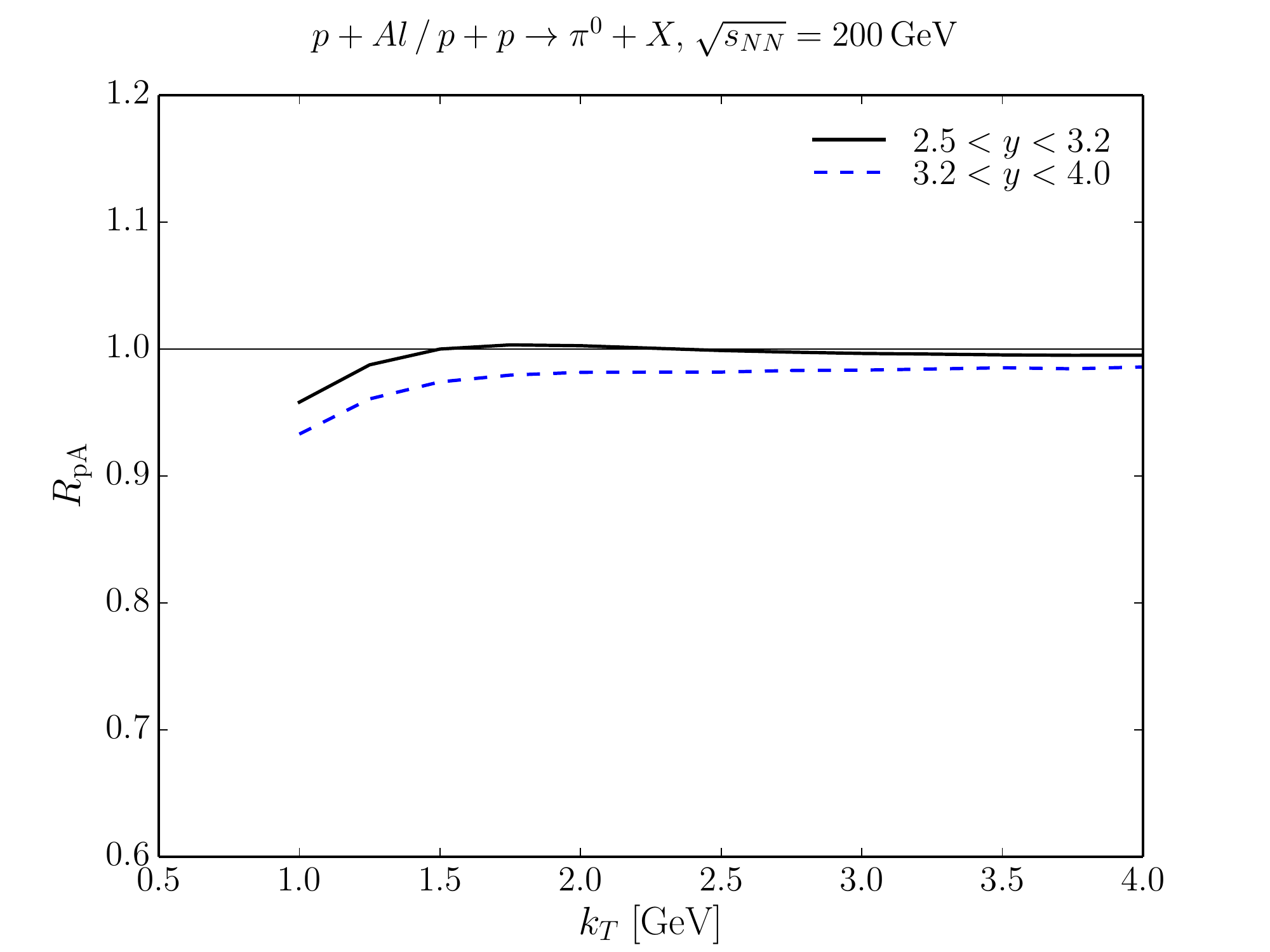} 
				\caption{Nuclear suppression factor for $\pi^0$ production in proton-aluminum collisions at $\sqrt{s_{NN}}=200$ GeV. }
		\label{fig:rpa_pion_200gev_al}
\end{figure}

\begin{figure}[tb]
\centering
		\includegraphics[width=0.5\textwidth]{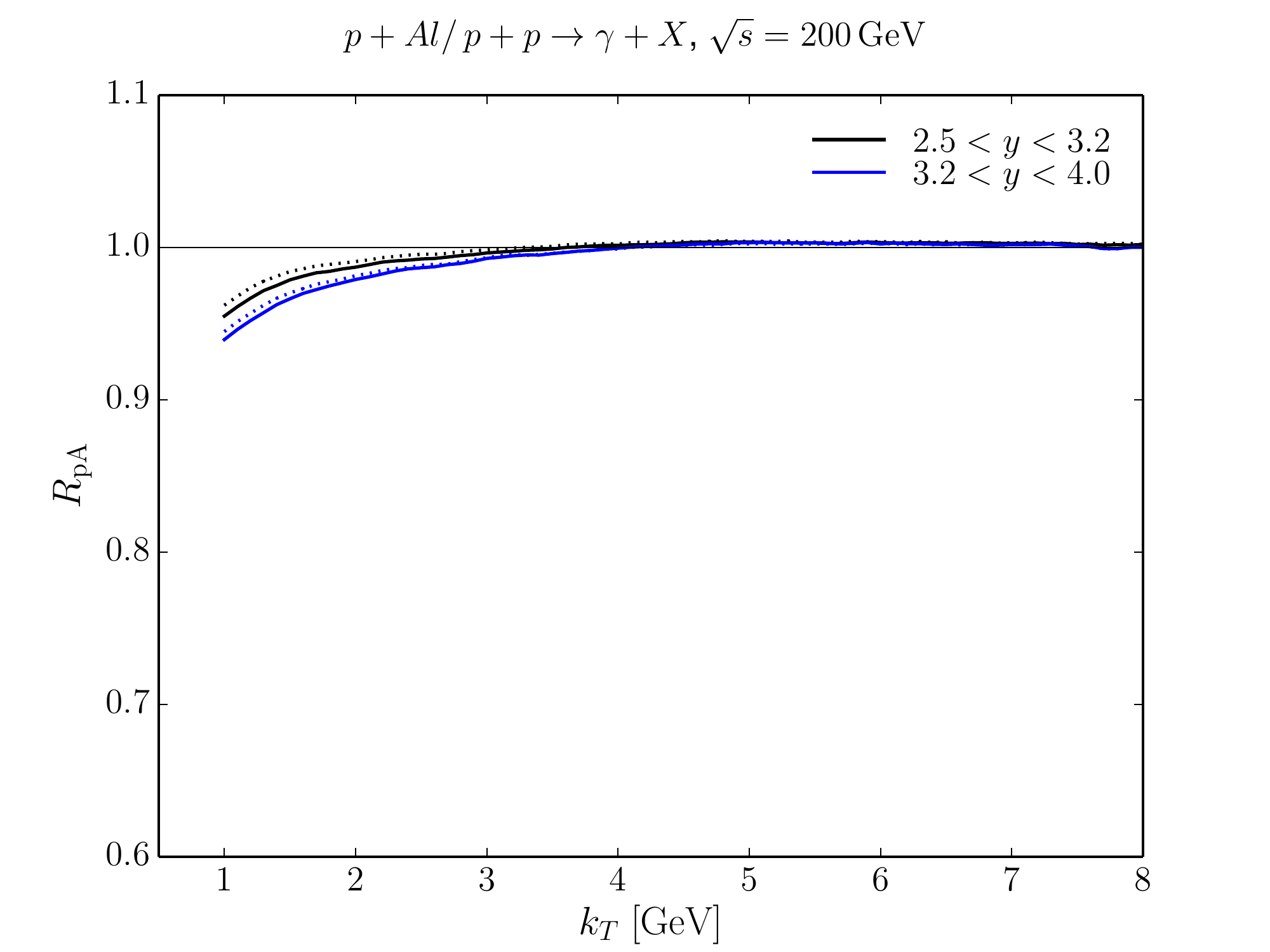} 
				\caption{Nuclear suppression factor for inclusive photon production in proton-aluminum collisions at $\sqrt{s_{NN}}=200$ GeV. Solid lines have isolation cut $R=0.4$ and dotted lines $R=0.1$. }
		\label{fig:rpa_photon_200gev_al}
\end{figure}

The nuclear suppression factor for $\pi^0$ production at forward rapidities is shown in Fig.~\ref{fig:rpa_pion_200gev_al}. As expected, we get basically no nuclear suppression, and the Cronin peak is practically invisible. Similarly, in the case of isolated photons for which $R_{pA}$ is shown in Fig.~\ref{fig:rpa_photon_200gev_al}, we do not expect any visible suppression at RHIC energies.

\section{Conclusions}
In this work we presented predictions for the nuclear modification factor in forward pion and direct photon production at RHIC and the LHC. The nuclear modification in our calculation is a result of the presence of strong saturation effects in the heavy nuclei at small $x$. We find that a significant suppression should be observed at moderate transverse momentum, and that the suppression grows strongly as a function of rapidity. We also expect that a  Cronin enhancement is seen at RHIC, in particular for pion production, and that it disappears when moving to LHC energies or to more forward rapidities.

In our framework the only input besides standard nuclear geometry comes from HERA deep inelastic scattering data, where the rcBK evolved dipole amplitude is fitted. In particular, in contrast to many other works, we do not introduce any additional parameters to control the saturation scale of the nucleus. Therefore, the nuclear modification factor is a robust observable, and we expect that the comparison of our results with future measurements at RHIC and the LHC will help to better understand the behavior of gluon densities at small $x$.

 \begin{figure}[tb]
\centering
		\includegraphics[width=0.5\textwidth]{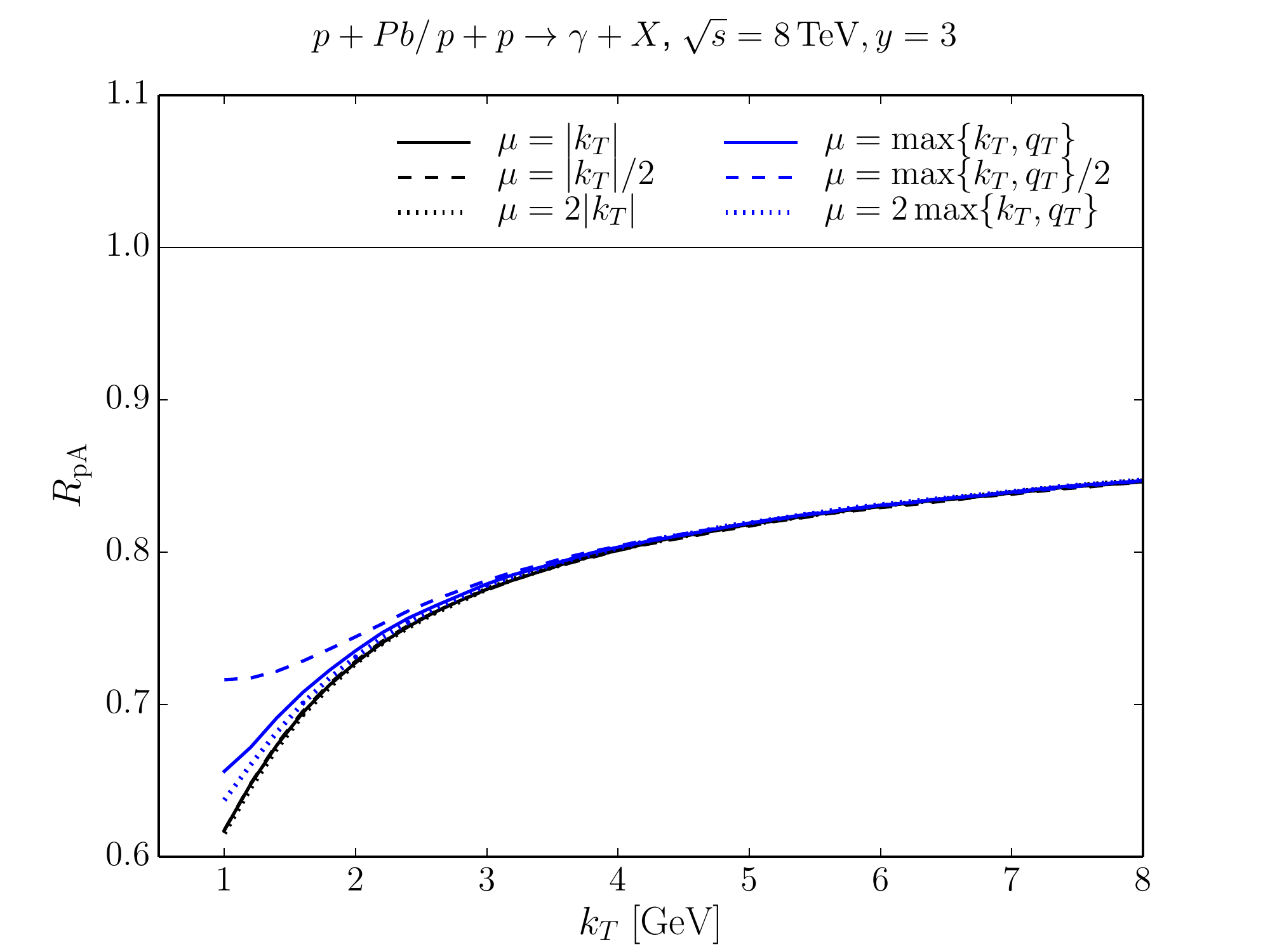} 
				\caption{Dependence on the scale at which parton distribution functions are evaluated when calculating photon production cross sections at the LHC. Here $q_T$ refers to the transverse momentum of the produced quark, and the photon rapidity is $y=3$. }
		\label{fig:rpa_photon_scale}
\end{figure}

\section*{Acknowledgments}
We thank E. Aschenauer and T. Peitzmann for discussions and I. Helenius for useful  comparisons. 
H.M is supported under DOE Contract No. DE-SC0012704. H.M. wishes to thank the University of Jyväskylä for hospitality and the Eemil Aaltonen Foundation for supporting his travel to collaborate with the University of Jyväskylä group. 
The work  of T.L. and B.D. has been supported by the Academy of Finland, projects 267321 and 303756, and by the European Research Council, grant
ERC-2015-CoG-681707.

 \appendix
 
 \section{Uncertainty related to the scale choice}
 \label{appendix:scale}
The leading order calculation does not set the scale at which the parton distributions and fragmentation functions should be evaluated when calculating cross sections. This introduces an uncertainty in our calculations.

The sensitivity of our results on the scale choice in the case of isolated photon production is demonstrated in Fig.~\ref{fig:rpa_photon_scale}. We find that except at very low transverse momentum, the scale uncertainty completely cancels in the nuclear modification factor. We note that the CTEQ parton distribution functions used in this work are only available at scales larger than $1.3\gev$, and at lower scales one introduces additional extrapolation uncertainties, which makes especially $|k_T|/2$ and $\max\{k_T,q_T\}/2$ results in Fig.~\ref{fig:rpa_photon_scale} unreliable at low $|k_T|$. We note that the scale variation likely underestimates the NLO corrections that originate, for example, from a) going beyond $1\to 2$ kinematics, and b) having different anomalous dimension due to the NLO evolution in the dipole amplitude which potentially affects the $p_T$ spectra~\cite{Lappi:2016fmu}.

\bibliographystyle{JHEP-2modlong.bst}
 \bibliography{../../refs}
 
\end{document}